\documentclass[preprint,showpacs,preprintnumbers,amsmath,amssymb]{revtex4}

\usepackage{graphicx}% Include figure files
\usepackage{dcolumn}
\usepackage{bm}
\usepackage{amsmath}
\usepackage{amsfonts}
\usepackage{amssymb}%
\usepackage{color}
\usepackage{orcidlink}
\usepackage[normalem]{ulem}
\begin{document}

\title{Creation of an inflationary epoch from an Emergent Universe through quantum tunneling}

%\author
\author{Eduardo I. Guendelman \orcidlink{0000-0003-1044-8055}}
\email{guendel@bgu.ac.il} \affiliation{Physics Department, Ben
Gurion University of the Negev, Beer Sheva 84105, Israel\\
Frankfurt Institute for Advanced Studies, Giersch Science Center, Campus Riedberg, Frankfurt am Main, Germany \\
Bahamas Advanced Studies Institute and Conferences,  4A Ocean Heights, Hill View Circle, Stella Maris, Long Island, The Bahamas}
\author{Ram\'on Herrera \orcidlink{0000-0002-6841-1629}}
\email{ramon.herrera@pucv.cl} \affiliation{ Instituto de
F\'{\i}sica, Pontificia Universidad Cat\'{o}lica de
Valpara\'{\i}so,  Avenida Brasil 2950, Casilla 4059,
Valpara\'{\i}so, Chile.}
\author{Pedro Labra$\tilde{n}$a \orcidlink{0000-0002-6390-8580}}
\email{plabrana@ubiobio.cl} \affiliation{Departamento de F\'{i}sica, Universidad del B\'{i}o-B\'{i}o, Casilla 5-C, Concepci\'on, Chile and
Centro de Ciencias Exactas, Universidad del B\'{i}o-B\'{i}o, Casilla 447, Chill\'an, Chile.}

\date{\today}% It is always \today, today,

\begin{abstract}
We consider that the very early Universe was well described by an Emergent Universe, whose geometry was that of a static and classically stable Einstein Universe, which is possible in certain Two Measures Theories (TMT). These solutions do not have a big bang, but rather they extend to arbitrarily early times. The TMT we consider are theories with spontaneously broken scale invariance and contain a dilaton field, crucial for the implementation of the scale invariance, that after the symmetry breaking acquires a nontrivial effective potential, which we study in the Einstein frame. Using the conserved quantity associated with the dilaton, we recast the cosmological evolution in terms of an effective potential $V(\dot{\phi})$, which exhibits a divergent barrier separating the static solution from the subsequent expanding evolution. Quantizing the corresponding minisuperspace Hamiltonian, we evaluate the tunneling probability in the WKB approximation. The tunneling action is well defined and finite once the branch of the square root is fixed by requiring a consistent semiclassical interpretation of the tunneling probability. We find that the barrier gives rise to a logarithmic contribution to the tunneling exponent, yielding a power-law rather than the usual exponential dependence of the tunneling probability, which for the parameter values considered is close to unity. The Universe thus emerges with a finite scale factor into a superinflationary phase that develops into slow-roll inflation. This quantum creation of an inflationary Universe does not represent a quantum creation of spacetime, which exists before the tunneling.

\end{abstract}

   \pacs{98.80.Cq, 04.20.Cv, 95.36.+x}% PACS, the Physics and Astronomy
\maketitle

\section{Introduction}

The question of the origin of the Universe, and of whether it had a beginning
at all, remains one of the deepest open problems in cosmology. The inflationary
paradigm \cite{Starobinsky:1980te, Guth:1981, Linde:1983gd} accounts for the
homogeneity, isotropy, and flatness of the observable Universe, as well as for
the spectrum of primordial perturbations, and is widely regarded as a successful
description of its very early stages \cite{Mukhanov:1981xt,Hawking:1982cz,Guth:1982ec}. Inflation by itself, however, does not
address what preceded it: under fairly general assumptions, classical
inflationary cosmologies are geodesically past-incomplete
\cite{BordeVilenkin:1994, BordeVilenkin:1996, Borde:2001nh}, and the singularity theorems of
general relativity imply that a classical expanding Universe is generically
traced back to an initial singularity \cite{PenroseHawking:1970, HawkingEllis:1973},
where the classical description of spacetime and matter breaks down.

A natural way to confront the initial singularity is to regard the birth of the
Universe as a quantum event. In the canonical approach to quantum cosmology, the
quantization of the Hamiltonian constraint of general relativity leads to the
Wheeler--DeWitt (WDW) equation \cite{DeWitt:1967yk}, ${\cal H}\,\Psi = 0$, whose
solution $\Psi$, the wave function of the Universe, is defined on superspace.
In practice one restricts to a minisuperspace spanned by the scale factor of a
homogeneous, isotropic and closed Friedmann--Robertson--Walker (FRW) geometry
together with a few matter variables, in which the WDW equation reduces to a
one-dimensional Schr\"odinger-like problem governed by an effective potential.
Unlike in ordinary quantum mechanics, the WDW equation must be supplemented by a
boundary condition on $\Psi$, which plays the role of a law for the initial state
and is not fixed by the dynamics itself. Two influential proposals have shaped
the field: the no-boundary proposal of Hartle and Hawking \cite{HartleHawking:1983}
and the tunneling proposal of Vilenkin \cite{Vilenkin:1982de, Vilenkin:1987kf}.
Both describe the quantum creation of a closed Universe ``from nothing'', and,
although they have been the subject of a long-standing debate as to which
boundary condition is the correct one, they share the essential feature of
describing a genuine quantum creation of spacetime itself.

In Vilenkin's tunneling scenario, a closed FRW Universe driven by a positive
vacuum energy is created by quantum nucleation through the barrier of the
effective potential $U(a)$. The process starts at vanishing scale factor,
$a = 0$, where no classical spacetime exists, and ends at the classical turning
point, beyond which the Universe expands and inflates. The expression
``creation from nothing'' thus refers to the absence of any classical spacetime or matter prior to nucleation: the Universe literally begins at zero size.

A conceptually different route to a non-singular beginning is provided by the Emergent Universe (EU) scenario \cite{EllisMaartens:2004, EllisMuruganTsagas:2004, Mulryne:2005ef}. Here the closed Universe is past-eternal: rather than originating in a singularity, it spends an indefinitely long time in a static Einstein state of finite radius before evolving towards an inflationary phase, so that the cosmological singularity is altogether absent and the "beginning of size" is replaced by a finite, non-zero seed. Since the Einstein static solution is unstable in general relativity, a considerable effort has been devoted to identifying frameworks in which it becomes stable, including modified gravity, braneworld and quantum-gravity corrections \cite{Mukherjee:2005zt, Mukherjee:2006ds,delCampo:2007mp, Banerjee:2007sg, Paul:2009csp, delCampo:2009kp, Wu:2009ah, Paul:2010jb, Paul:2011nw, Labrana:2011np, Chattopadhyay:2011fp, Liu:2012ww, Cai:2012yf, Chakraborty:2014ora, Alesci:2016xqa, Labrana:2018bkw, Labrana:2021zyy, Bengochea:2021jvt, Bhuyan:2024xnl}, as well as to the observational signatures of the emergent phase \cite{Labrana:2013oca,Rios:2016trs,Labrana:2016jmm,Martineau:2018isp,Huang:2022mxj,Palermo:2022dim,Huang:2022hye, Huang:2023qst,Huang:2026dlm}.
%
%Various realizations of this scenario have also been developed by us in Refs.~\cite{delCampo:2007mp, delCampo:2009kp, Labrana:2011np, Labrana:2018bkw, Labrana:2021zyy, Shabani:2024ler}. 
%
Concrete realizations of this idea have been constructed, in particular, within scale-invariant  TMT \cite{delCampo:2010kf,delCampo:2011mq, Guendelman:2013dka, delCampo:2015yfa, Guendelman:2014bva,Guendelman:2015uca, Guendelman:2023jsk}, where a static and classically stable Einstein Universe arises naturally.
A recurrent difficulty of these scenarios concerns the stability of the static state against quantum decay, since the Einstein static Universe can in principle collapse via quantum tunneling, which makes the required past-eternal phase delicate to sustain \cite{Mithani:2011en, Mithani:2014jva}.

These two lines of thought, quantum creation by tunneling and the emergent
Universe, suggest a natural question. In the standard tunneling scenario the
Universe is born from nothing, at $a = 0$, whereas in the emergent scenario it
rests instead in a static configuration of finite radius. Is it then possible to
formulate a quantum-cosmological creation mechanism in which the Universe does
not tunnel from $a = 0$, but rather from a pre-existing classical state
of finite, constant scale factor, namely the radius of the static Einstein
(emergent) Universe? In the present work we address precisely this problem.
Working within the framework of a scale-invariant Two Measures Theory, we
consider a Universe that is initially in the classical emergent (static
Einstein) state, of finite radius, and show that it can be quantum mechanically
destabilized, tunneling not from nothing but from this finite radius into an
expanding, superinflationary phase that subsequently develops into a slow-roll
inflationary stage. To this end, we reconstruct the effective potential as a function of $\dot{\phi}$, analyze the WKB tunneling action, and evaluate the corresponding tunneling probability within the TMT framework. A distinctive feature of this construction is that spacetime
already exists prior to the tunneling event, so that, unlike in the
Hartle--Hawking or Vilenkin scenarios, the process does not describe a quantum
creation of spacetime.

As a direct consequence, the ambiguity in the choice of boundary conditions for
the wave function of the Universe, which lies at the core of the long-standing
debate between the followers of the Hartle--Hawking and the Vilenkin
proposals, does not arise in our construction: the relevant initial state is
known and is given by the classical Emergent Universe solution. A complementary
realization of the same scenario, in which the Einstein static state is instead
supported by a tachyonic field together with a Casimir energy density, is
developed in a companion work \cite{tachyon:companion}.

The paper is organized as follows. In Section~\ref{R1} we introduce the Two Measures
Theory with spontaneously broken scale invariance that we shall use. In
Section~\ref{EU} we present the Emergent Universe solution and derive the effective
potential for $\dot{\phi}$, where $\phi$ is the dilaton field. In Section~\ref{Tun} we study the
quantum tunneling from the Emergent Universe solution to the inflationary phase.
Section~\ref{DC} is devoted to a general discussion of the implications of this
work. Finally, in Appendix~\ref{app2} we present the three solutions for $\dot{\phi}$ as a function of the scale factor. In this work, we choose units such that $c=\hbar=1$.

\section{Two Measures Theory}\label{R1}

In this section  we  will present a brief analysis of the TMT \cite{Guendelman:1996qy,Guendelman:1999tb,Guendelman:2002js}. 
The action associated with the TMT can be written as \cite{Guendelman:1999tb,Guendelman:2002js}
\begin{equation}
    S_{TMT} = \int L_{1}\Phi d^{4}x +\int L_{2}\sqrt{-g}d^{4}x\,,
\label{S}
\end{equation}
where $L_{1}$ and $L_{2}$ are two Lagrangians and  the quantities
$\sqrt{-g}$ and $\Phi$ are
 two measures of
integration. Here, the quantity $\sqrt{-g}$ is 
the usual measure of
integration  in the 4-dimensional spacetime manifold
related with the metric
 $g_{\mu\nu}$ and $\Phi$
   is the new measure of integration  in the same
 manifold. In addition, the scalar-density measure  $\Phi$ becomes a  total derivative  and it is 
 defined by means of  four scalar fields $\varphi_{a}$
 with $a=1,2,3,4$, (see Refs.~\cite{Guendelman:2000mt,Guendelman:2002fb})
\begin{equation}
\Phi
=\varepsilon^{\mu\nu\alpha\beta}\varepsilon_{abcd}\partial_{\mu}\varphi_{a}
\partial_{\nu}\varphi_{b}\partial_{\alpha}\varphi_{c}
\partial_{\beta}\varphi_{d}.
\label{Phi}
\end{equation}

The Lagrangian densities $L_{1}$ and $L_{2}$ are assumed to be
functions of all matter fields, the dilaton field, the metric, the
connection, but independent of the measure fields
  $\varphi_{a}$. When the measure fields enter the theory exclusively through the measure $\Phi$, the action associated with the TMT given by Eq.~(\ref{S}) exhibits an infinite-dimensional symmetry. For the specific realization defined by Eq.~(\ref{Phi}), this symmetry  is generated by transformations of the form $\varphi_{a}\rightarrow\varphi_{a}+f_{a}(L_{1})$, where
$f_{a}(L_{1})$ are arbitrary functions of  $L_{1}$ \cite{Guendelman:1999tb}. Also, within this framework, we consider that all fields, including the metric, the connection, and the measure fields, constitute independent variables.
  
By varying the measure fields $\varphi_{a}$, one obtains
$B^{\mu}_{a}\partial_{\mu}L_{1}=0 $, where the quantity $B_a^{\mu}$ is defined as 
$B^{\mu}_{a}=\varepsilon^{\mu\nu\alpha\beta}\varepsilon_{abcd}
\partial_{\nu}\varphi_{b}\partial_{\alpha}\varphi_{c}
\partial_{\beta}\varphi_{d}.\label{varphiB}$
Hence, since $\det (B^{\mu}_{a}) = \frac{4^{-4}}{4!}\Phi^{3}$ it follows
that, for  $\Phi\neq 0$,
\begin{equation}
 L_{1}=sM^{4} =\mbox{constant},
\label{varphi}
\end{equation}
where the parameter $s=\pm 1$ and the quantity $M$ is  an  integration constant  with 
dimension of mass.

Within  the
context of TMT, a  dilaton field $\phi$ enables the spontaneous breaking of the global scale invariance. We assume
that the theory is invariant under the global scale transformations given by 
\begin{equation}
    g_{\mu\nu}\rightarrow e^{\theta }g_{\mu\nu}, \quad
\Gamma^{\mu}_{\alpha\beta}\rightarrow \Gamma^{\mu}_{\alpha\beta},
\quad \varphi_{a}\rightarrow \lambda_{ab}\varphi_{b}\quad
\text{where} \quad \det(\lambda_{ab})=e^{2\theta}, \quad
\phi\rightarrow \phi-\frac{M_{p}}{\alpha}\theta . \label{st}
\end{equation}

Also, we adopt an action that, apart from the structural modifications implied by TMT, contains no exotic field or interactions and closely follows the conventional minimal coupled scalar gravity framework. In this way, following Ref.~\cite{Guendelman:2006af} we retain the general structure of the action defined by Eq.~(\ref{S}), obtaining

\begin{eqnarray}
S_{TMT} &=&\int d^{4}x e^{\alpha\phi /M_{p}}
\Big[-\frac{1}{2\,\kappa}R(\Gamma ,g)(\Phi +b_{g}\sqrt{-g})+(\Phi
+b_{\phi}\sqrt{-g})\frac{1}{2}g^{\mu\nu}\phi_{,\mu}\phi_{,\nu}
\label{totaction}
\\
&-& e^{\alpha\phi /M_{p}}\left(\Phi V_{1}
+\sqrt{-g}V_{2}\right)\Big],\nonumber
\end{eqnarray}
where the parameter $\kappa =8\pi/M_p^2 $ with $M_p$ denoting  the four-dimensional
Planck mass. In addition, the quantities $V_1$ and $V_2$ are two constants  and $b_g$ and $b_\phi$ are two dimensionless real parameters \cite{Guendelman:2006af}. 

Considering a conformal metric $\tilde{g}_{\mu\nu}$ in  the ``Einstein frame'' given by
\begin{equation}
\tilde{g}_{\mu\nu}=e^{\alpha\phi/M_{p}}(\zeta +b_{g})g_{\mu\nu},
\label{ct}
\end{equation}
where the quantity $\zeta$ is defined as $\zeta \equiv\frac{\Phi}{\sqrt{-g}}$, the action given by Eq.~(\ref{totaction}) can be reduced to the standard General Relativity (GR) action. Here we note that  the
conformal metric $\tilde{g}_{\mu\nu}$  is invariant under the scale transformations given by
Eq.~(\ref{st}). Thus, assuming the change of variables  to the Einstein frame, the
gravitational equations assume their  standard GR form \cite{Guendelman:2006af}
\begin{equation}
G_{\mu\nu}(\tilde{g}_{\alpha\beta})=\kappa\,T_{\mu\nu}^{eff},
 \label{gef}
\end{equation}
where    $G_{\mu\nu}(\tilde{g}_{\alpha\beta})$ is the Einstein tensor.

The effective energy-momentum tensor in the Einstein frame $T_{\mu\nu}^{eff}$ is given by  
\begin{eqnarray}
T_{\mu\nu}^{eff}&=&\frac{\zeta +b_{\phi}}{\zeta +b_{g}}
\left(\phi_{,\mu}\phi_{,\nu}-\frac{1}{2}
\tilde{g}_{\mu\nu}\tilde{g}^{\alpha\beta}\phi_{,\alpha}\phi_{,\beta}\right)
-\tilde{g}_{\mu\nu}\frac{b_{g}-b_{\phi}}{2(\zeta +b_{g})}
\tilde{g}^{\alpha\beta}\phi_{,\alpha}\phi_{,\beta}
+\tilde{g}_{\mu\nu}V_{eff}(\phi;\zeta,M),
 \label{Tmn}
\end{eqnarray}
where the quantity $V_{eff}(\phi;\zeta,M)$ is given by 
\begin{equation}
V_{eff}(\phi;\zeta ,M)=
\frac{b_{g}\left[sM^{4}e^{-2\alpha\phi/M_{p}}+V_{1}\right]
-V_{2}}{(\zeta +b_{g})^{2}}. \label{Veff1}
\end{equation}

In addition, the scalar field  
  $\zeta$  is fixed by the consistency of
Eq.~(\ref{gef}) with Eq.~(\ref{varphi}), leading  to the constraint
\begin{eqnarray}
&&(b_{g}-\zeta)\left[sM^{4}e^{-2\alpha\phi/M_{p}}+
V_{1}\right]-2V_{2}-\delta\cdot b_{g}(\zeta +b_{g})Z
=0,\label{constraint2}
\end{eqnarray}
where the quantities $Z$ and $\delta$ are defined as
$Z\equiv\frac{1}{2}\tilde{g}^{\alpha\beta}\phi_{,\alpha}\phi_{,\beta}$
and $\delta =\frac{b_{g}-b_{\phi}}{b_{g}}$, respectively.

The effective energy-momentum tensor given by Eq.~(\ref{Tmn}) can be identified  
with that of a  perfect fluid $T_{\mu\nu}^{eff}=(\rho
+p)u_{\mu}u_{\nu}-p\tilde{g}_{\mu\nu}$, where the four-velocity 
is given by $u_{\mu}=\frac{\phi_{,\mu}}{(2Z)^{1/2}}$. Here, $\rho$ and $p$ are the energy density and pressure associated with the perfect fluid. Thus, considering the solution $\zeta =\zeta(\phi,Z;M)$
given by  Eq.~(\ref{constraint2}), the energy
density $\rho$ and pressure $p$ can be written as \cite{Guendelman:2006af}

\begin{equation}
\rho(\phi,Z;M) =Z+ \frac{(sM^{4}e^{-2\alpha\phi/M_{p}}+V_{1})^{2}-
2\delta b_{g}(sM^{4}e^{-2\alpha\phi/M_{p}}+V_{1})Z -3\delta^{2}
b_{g}^{2}Z^2}{4[b_{g}(sM^{4}e^{-2\alpha\phi/M_{p}}+V_{1})-V_{2}]},
\label{rho1}
\end{equation}
and
\begin{equation}
p(\phi,Z;M) =Z- \frac{\left(sM^{4}e^{-2\alpha\phi/M_{p}}+V_{1}+
\delta b_{g}Z\right)^2}
{4[b_{g}(sM^{4}e^{-2\alpha\phi/M_{p}}+V_{1})-V_{2}]}. \label{p1}
\end{equation}
The different constants associated with this model are subject to
the observational constraints and stability conditions given by 
Eqs.~(\ref{ES1})--(\ref{ES3}), as  studied in Ref.~\cite{delCampo:2011mq}.

\section{Emergent Universe}\label{EU}

The EU scenario is based on the assumption that the Universe originates from a past-eternal Einstein static (ES) state, subsequently evolving toward an inflationary phase and a hot Big Bang era. In a series of papers~\cite{delCampo:2010kf,delCampo:2011mq,Guendelman:2014bva,Guendelman:2013dka,Guendelman:2015uca,delCampo:2015yfa}, we have studied a class of EU scenarios based on a spontaneously broken scale invariance induced by the dynamics of a Two Measures Theory.

In the present work we focus in particular on the model of Ref.~\cite{delCampo:2011mq}, in which the ES solution is classically stable under homogeneous and isotropic perturbations and which, as shown in Ref.~\cite{delCampo:2015yfa}, does not suffer the semiclassical instability towards collapse reported by Vilenkin~\cite{Mithani:2011en}. We note that the results obtained in this case can also be applied to the models studied in Refs.~\cite{delCampo:2010kf,Guendelman:2014bva}, which present symmetries similar to those of the model of Ref.~\cite{delCampo:2011mq}.

We begin  by considering a metric describing a closed  FRW spacetime, for which  the line  element $ds^2$ is defined as  
\begin{equation}
ds^2 =dt^2 - a(t)^2 \left(\frac{dr^2}{1 -r^2}+ r^2(d\theta^2
+\sin^2\theta d\phi^2)\right),  \label{Fr}
\end{equation}
where $a(t)$ is  the scale factor and $t$ is the cosmic time. 

We further assume that the scalar field is homogeneous, i.e., 
$\phi$ depends  only on
 the cosmic time $t$, such that $\phi({\bf{x}},t)=\phi(t)$.

For the dynamics, we will assume a stage in which the scalar field $\phi$ is moving
in the extreme left region $\phi \rightarrow -\infty  $ of the effective potential given by Eq.~(\ref{Veff1}). In this
context, the expressions for the energy density $\rho$ and pressure $p$
given by Eqs.~(\ref{rho1}) and (\ref{p1}) can be written as
\begin{equation}\label{eq.density}
\rho = \frac{A}{2} \dot{\phi}^2 + 3B\dot{\phi}^4 + C,
\end{equation}
and
\begin{equation}
p = \frac{A}{2} \dot{\phi}^2 +B\dot{\phi}^4 - C,\label{presion}
\end{equation}
respectively. Here, the  values of 
  the constants $A$,  $B$ and $C$ are defined by
\begin{eqnarray}
A = 1- \frac{2\delta b_g V_1}{4(b_g V_1 - V_2)}\,,\,\,\,\, B =
-\frac{\delta^2 b^2_g }{4(b_g V_1 - V_2)}\,,\,\,\,\,
\mbox{and}\,\,\,\, C = \frac{ V^2_1}{4(b_g V_1 - V_2)}\,.\label{C}
\end{eqnarray}
In the following, the dots denote derivatives with respect to  cosmic time $t$.

As shown in Ref.~\cite{delCampo:2011mq}, the emergent Universe can evolve 
into an inflationary stage only if the constant $C>0$, i.e., if $b_gV_1>V_2$.

In analyzing the stability of the model of Ref.~\cite{delCampo:2011mq}, we noted that its Wheeler-DeWitt potential leaves open the possibility of additional tunneling channels, corresponding to a transition from the ES solution to a different value of the scale factor, from which the Universe would subsequently begin to evolve classically. It is precisely this possibility that we explore in the present work.

We now continue with the review of Ref.~\cite{delCampo:2011mq}, in which the constant value of the scale factor characterizing the Einstein static Universe, $a=a_+$, is given by
\begin{equation}\label{consta}
a_+ = \sqrt{\left(\frac{3}{8\pi G}\right)\frac{12B}{A^2 + 24B\,C -
A\sqrt{A^2 + 12B\,C}}}\,\,.
\end{equation}
In order for this static solution to be stable under homogeneous and isotropic perturbations, the following conditions must be satisfied, as discussed in Ref.~\cite{delCampo:2015yfa}:
\begin{eqnarray}
0.5<y<0.54\,,\,\,\,\,\,\,\,\,\,\,\,\,\,\,B<0\,,\,\,\,\,\,\,\,\,\,\,\,\,\,\,\,\mbox{and} \label{ES1}\\
-\frac{1}{64B} < C < -\sqrt{\frac{3}{B^2}} -
\frac{7}{4B}\,,\label{ES3}
\end{eqnarray}
where the constant $A$ is defined as $A=1-y$, with the quantity $y$ given by $y=\frac{2\delta b_g C}{V_1}$.

In addition, as established in  Ref.~\cite{delCampo:2011mq},  there is a conserved quantity
$\Pi_\phi$. Indeed, in the limit $\phi\rightarrow -\infty$, the theory acquires an additional symmetry under the shift $\phi\rightarrow \phi+ c$, with $c$ a constant. In the
Einstein frame, the action takes the form 
\begin{equation}
S_E=\frac{1}{\kappa}\left[\int\,
R\,\sqrt{-g}\,d^4x\,+\,\int\,p\,\sqrt{-g}\,d^4x\right],\label{Acc}
\end{equation}
and the symmetry $\phi\rightarrow \phi+ c$ leads to the conservation law \cite{delCampo:2011mq}

\begin{equation}\label{Cons1}
a^3(t)\,[A\dot{\phi}+4\,B\dot{\phi}^3]=\Pi_\phi=\mbox{constant}\,.
\end{equation}
Without loss of generality, we take $\Pi_\phi$ to be positive.
From Eq.~(\ref{Cons1}), we can obtain a relation between the scale factor and $\dot{\phi}$ given by

\begin{equation}\label{Cons1a}
a(\dot{\phi}) =
\left(\frac{\Pi_\phi}{A\dot{\phi}+4\,B\dot{\phi}^3}\right)^{1/3} \,.
\end{equation}

Because $\Pi_\phi
> 0$, the range of $\dot{\phi}$ is  given by  $-\infty < \dot{\phi}<
-\sqrt{\frac{A}{4|B|}}$ or alternatively  $0 <\dot{\phi}< \sqrt{\frac{A}{4|B|}}$.

In the first case, in which $-\infty < \dot{\phi}<
-\sqrt{\frac{A}{4|B|}}$, the scale factor as a function of  $\dot{\phi}$, i.e., $a(\dot{\phi})$, goes to zero when $\dot{\phi}\rightarrow-\infty$  and diverges when $\dot{\phi} \rightarrow
-\sqrt{\frac{A}{4|B|}}$. However, in this 
  region, the energy density $\rho$ defined by Eq.~(\ref{eq.density}) is negative, 
 which is therefore not of physical interest here.

Alternatively, in the range $0 <\dot{\phi}< \sqrt{\frac{A}{4|B|}}$, the scale factor 
$a(\dot{\phi})$ has an extremum (minimum) at $\dot{\phi} =
\dot{\phi}_0$,  where the scale factor  $a(\dot{\phi}=\dot{\phi}_0)=a_0$, and these values are given by 
\begin{eqnarray}
\dot{\phi}_0 = \sqrt{\frac{A}{12|B|}}\,,\,\,\,\,\,\,\mbox{and}\,\,\,\,\,\,\,\,\,
a_0 &=&
\left(\frac{12|B|}{A}\right)^{1/6}\left[\frac{3\Pi_{\phi}}{2A}\right]^{1/3}\,,\label{a0}
\end{eqnarray}
respectively. In addition, we note that 
 from Eq.~(\ref{Cons1a}), the scale factor $a(\dot{\phi})$
diverges when $\dot{\phi}\rightarrow 0$  or when 
$\dot{\phi}\rightarrow\sqrt{\frac{A}{4|B|}}$.

\begin{figure}[h]
\begin{center}
\includegraphics[width=2.3in,angle=0,clip=true]{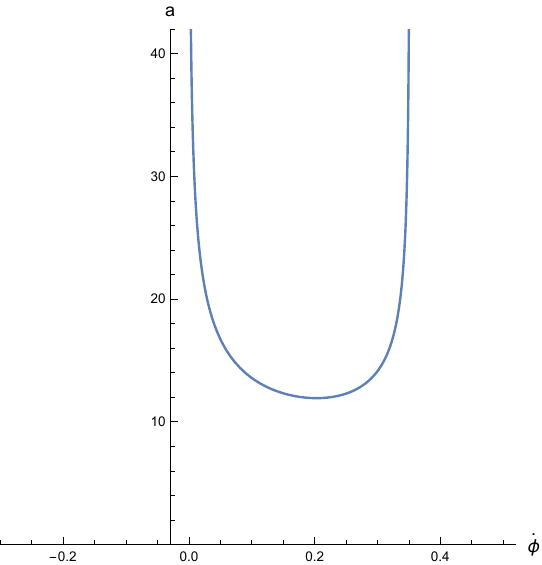}
\caption{Scale factor $a$ as a function 
of $\dot{\phi}$, obtained from Eq.~(\ref{Cons1}). Here we have considered the values  $B=-1$, $C=0.016$, $y
=0.505964$ and $\Pi_\phi = 113.41$.\label{Tres-Sol}}
\end{center}
\end{figure}

Moreover, we  note that values of the  scale factor smaller than $a_0$, i.e., $a(\dot{\phi})<a_0$, lie outside the domain  where the physical
solutions exist. To illustrate  this,  Fig.~\ref{Tres-Sol} shows the scale factor as a function of  
$\dot{\phi}$, for the values 
$B=-1$, $C=0.016$, $y
=0.505964$ and $\Pi_\phi = 113.41$.

In Appendix~\ref{app2} we present the inverse relation of Eq.~(\ref{Cons1a}), i.e., $\dot{\phi}(a)$, which yields three solutions for $\dot{\phi}$ in terms of the scale factor, given by Eqs.~(\ref{cu1}), (\ref{cu2}) and (\ref{cu3}). These solutions are plotted in Fig.~\ref{F3}, using $\Pi_\phi=113.41$ together with the parameter values of Ref.~\cite{delCampo:2011mq}.

%%%%%%%%%%%%%%%%%%%%%%%%%
%%%%%%%%%%%%%%%%%%%%%%%%%%%%%%%

The classical theory which describes the dynamics of this Universe is governed by the Friedmann equation. From Eqs.~(\ref{Fr}) and (\ref{Acc}), for a closed Universe with energy density $\rho$, it can be written as
\begin{equation}\label{Friedmann}
\left(\frac{\dot{a}}{a}\right)^2 = H^2=\frac{8\pi G}{3}\,\rho(a) -
\frac{1}{a^2}\,,
\end{equation}
where $H=\dot{a}/a$ is the Hubble parameter. 

Additionally,  the equation for $\dot{H}$ for a closed Universe is given by

\begin{equation}\label{DH}
\dot{H} = - 4\pi\,G(\rho + p) + \frac{1}{a^2}\,.
\end{equation}

Now,  using the relation between the scale factor and $\dot{\phi}$ defined by  Eq.~(\ref{Cons1}), we can rewrite the Friedmann equation given by 
Eq.~(\ref{Friedmann})  in terms of $\dot{\phi}$,  becoming 
\begin{equation}
\ddot{\phi}^2 + V(\dot{\phi}) = 0\,,\label{ddf}
\end{equation}
where the quantity $V(\dot{\phi})$ can be interpreted as an effective potential, depending on $\dot{\phi}$ as   
\begin{equation}
V(\dot{\phi}) = \frac{9(A\dot{\phi} + 4B\dot{\phi}^3)^2}{(A +
12B\dot{\phi}^2)^2}\,\left[ \frac{1}{\Pi_\phi^{2/3}}(A\dot{\phi} +
4B\dot{\phi}^3)^{2/3} -
\frac{\kappa}{3}\left(\frac{A}{2}\dot{\phi}^2 + 3B\dot{\phi}^4
+C\right) \right].\label{V1}
\end{equation}

\begin{figure}[t]
\begin{center}
\includegraphics[width=4.5in,angle=0,clip=true]{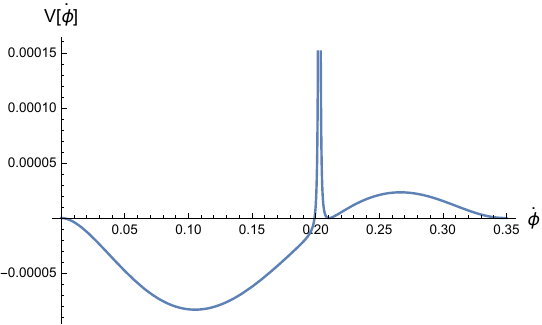}
\caption{Potential $V(\dot{\phi})$ as a function of $\dot{\phi}$, for  $\dot{\phi}
>0$. Here, the infinite barrier is located at $\dot{\phi}_0=0.20$.\label{FPP1}}
\end{center}
\end{figure}

\begin{figure}[t]
\begin{center}
\includegraphics[width=3.5in,angle=0,clip=true]{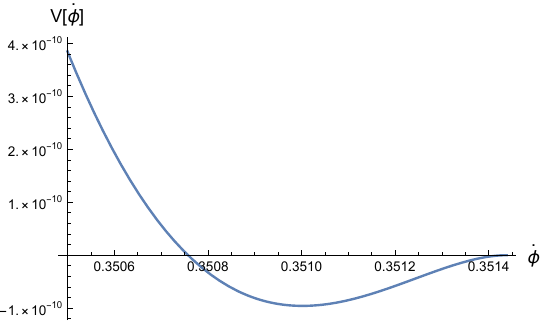}
\caption{Potential $V(\dot{\phi})$ as a function of $\dot{\phi}$, to the right of the finite barrier approaching the maximum value $\sqrt{\frac{A}{4|B|}}$.}\label{FPP2}
\end{center}
\end{figure}

As an example, Fig.~\ref{FPP1} shows the potential $V(\dot{\phi})$, where we have considered the value of the constant $\Pi_\phi = 113.41$ together with the parameter values used in Ref.~\cite{delCampo:2011mq}, $B=-1$, $C=0.016$ and $y=0.505964$. From Eq.~(\ref{ddf}) we have $\ddot{\phi}^2 = -V(\dot{\phi})$, so that the classically allowed regions are those in which $V(\dot{\phi})\leq 0$, whereas the regions with $V(\dot{\phi})>0$ are classically forbidden and can only be traversed by tunneling. In Fig.~\ref{FPP1} we can identify a potential well with $V(\dot{\phi})<0$ extending from $\dot{\phi}=0$ up to the divergent barrier located at $\dot{\phi} = \dot{\phi}_0 = \sqrt{A/(12|B|)}\simeq0.20$. Immediately to the right of this barrier the potential vanishes again, at the value
$\dot{\phi}_+=\sqrt{(-A+\sqrt{A^2+12BC})/(6B)}\simeq0.21$, which corresponds to the static (ES) Universe. Beyond it, the potential exhibits a second, finite barrier, on the far side of which it becomes negative again, defining a second classically allowed region that extends up to $\dot{\phi}=\sqrt{A/(4|B|)}$, see Fig.~\ref{FPP2}.

 From this potential,  we  note that the possibility of
tunneling through the divergent barrier located at $\dot{\phi}_0=\sqrt{A/(12|B|)}$  is allowed, see Ref.~\cite{Guendelman:2025swp}.
%, see 
%\cite{Dittrich:1985riz}.%,Dittrich:1%984cf}. 
Thus, the tunneling process proceeds from the 
static value $\dot{\phi}_+$ to the value $\dot{\phi}_-$
 situated to the left of the infinite barrier, where the potential also vanishes, i.e., $V(\dot{\phi}_-)=0$, see Fig.~\ref{FPP1}.

As discussed in Ref.~\cite{delCampo:2011mq}, this tunneling process does not represent a collapse of the Universe towards $a\rightarrow0$, but rather a transition from the static configuration, with scale factor $a_+$ given by Eq.~(\ref{consta}), to the configuration with $a_-=a(\dot{\phi}_-)$, from which the Universe subsequently begins to evolve. Indeed, after the tunneling event the Universe emerges with $a(\dot{\phi}_-)=a_-$ and $\dot{a}_-=0$, see Eqs.~(\ref{Friedmann}) and (\ref{V1}). Then, as $\dot{\phi}$ goes asymptotically to zero after tunneling, see Fig.~\ref{FPP1}, the scale factor grows, see Fig.~\ref{Tres-Sol}. During this process, the Hubble parameter goes from zero to a constant value $H_0$. This evolution will be discussed in more detail in Sec.~\ref{Tun}.

We note that there is also the possibility of tunneling through the finite barrier 
located to the right of the potential $V(\dot{\phi})$, into the second classically allowed region discussed above, see Fig.~\ref{FPP2}. After tunneling through this branch, 
$\dot{\phi}$ evolves within this region towards $\sqrt{A/(4|B|)}$, where the Universe reaches 
a de Sitter solution with $H= \sqrt{\frac{\kappa(A^2+16BC)}{48B}}$, since in this limit $\rho + p = 0$, see Ref.~\cite{delCampo:2011mq}. However, this case does not 
correspond to a standard slow-roll inflationary regime, as the kinetic term in Eq.~(\ref{eq.density}) 
is no longer negligible. In this paper, we restrict our analysis to the tunneling through the infinite barrier, and leave a detailed investigation of this scenario for future work.

%%%%%%%%%%%%%%%%%%%%%
%%%%%%%%%%%%%%%%%%%%%%%%%%%%%%%

%%%%%%%%%%%%%%%%%%%%%
%%%%%%%%%%%%%%%%%%%%%%%%%%%%%%%

\section{Tunneling from EU to Inflation}\label{Tun}

In this section, we study the tunneling of $\dot{\phi}$ through the infinite barrier of the effective potential, from the static value $\dot{\phi}_+$ to the value $\dot{\phi}_-$ located to its left, see Fig.~\ref{FPP1}. To determine the tunneling probability, we first reconstruct the action associated with the effective potential defined by Eq.~(\ref{V1}). Equation~(\ref{ddf}) can also be obtained from an effective action, as follows: first, we write the effective action associated to the scalar field in terms of the Lagrangian density $L$ as 
\begin{equation}
S=\int\,L(\dot{\phi},\ddot{\phi}) \,dt=\int\,L \,dt=\int\,dt\,[\ddot{\phi}^2-V(\dot{\phi})].
\end{equation}

By varying this action and integrating by parts, one obtains
\begin{equation}
\delta S=\int\,dt\,\left(\frac{\partial L}{\partial \dot{\phi}}-\frac{d}{dt}\left(\frac{\partial L}{\partial\ddot{\phi}}\right)\right)\delta\dot{\phi}.
\end{equation}
Integrating by parts once more, the variation of the action can be written in terms of $\delta\phi$ as
\begin{equation}
\delta S=-\int\,dt\,\frac{d}{dt}\left(\frac{\partial L}{\partial \dot{\phi}}-\frac{d}{dt}\left(\frac{\partial L}{\partial\ddot{\phi}}\right)\right)\delta{\phi}=0,
\end{equation}
which implies that 
\begin{equation}
\frac{d}{dt}\left(\frac{\partial L}{\partial \dot{\phi}}-\frac{d}{dt}\left(\frac{\partial L}{\partial\ddot{\phi}}\right)\right)=0.\label{Ds}
\end{equation}
Thus, the first integral of Eq.~(\ref{Ds}) becomes
\begin{equation}
\frac{\partial L}{\partial \dot{\phi}}-\frac{d}{dt}\left(\frac{\partial L}{\partial\ddot{\phi}}\right)=M=\mbox{const.},\label{M}
\end{equation}
where the constant $M$ can be absorbed by defining the effective Lagrangian $L_{eff}(\dot{\phi},\ddot{\phi})$ as 
\begin{equation}
L_{eff}(\dot{\phi},\ddot{\phi})=L_{eff}=L-M\dot{\phi},
\end{equation}
and Eq.~(\ref{M}) then reduces to 
\begin{equation}
\frac{\partial L_{eff}}{\partial \dot{\phi}}-\frac{d}{dt}\left(\frac{\partial L_{eff}}{\partial\ddot{\phi}}\right)=0.\label{M2}
\end{equation}

From Eq.~(\ref{ddf}) we identify that the effective Lagrangian
can be written as 
\begin{equation}
L_{eff}=\ddot{\phi}^2 - V_{eff}(\dot{\phi}) \,,\label{ddf1}
\end{equation}
where the new effective potential $V_{eff}(\dot{\phi})$ is defined as
\begin{equation}
V_{eff}(\dot{\phi})=V(\dot{\phi})+M\dot{\phi}.
\end{equation}
Thus, from Eq.~(\ref{M2}), the equation of motion becomes
\begin{equation}
2\dddot{\phi}+\frac{\partial V}{\partial\dot{\phi}}+M=0.
\end{equation}
By imposing that the integration constant $M=0$, we recover the Friedmann constraint given by Eq.~(\ref{ddf}). Under this condition,  the Hamiltonian associated with  our emergent Universe can be written as
\begin{equation}
{\cal {H}}(\dot{\phi},p_{\dot{\phi}})={\cal{H}}=\frac{p_{\dot{\phi}}^2}{4}+V(\dot{\phi}),
\end{equation}
where the momentum $p_{\dot{\phi}}$ is given by
\begin{equation}
p_{\dot{\phi}}= \frac{\partial L}{\partial\ddot{\phi}}=2\ddot{\phi},
\end{equation}
where $V(\dot{\phi})$ is defined by Eq.~(\ref{V1}).

Within the framework  of quantum theory, the Universe can be described by
a wave function $\psi(\dot{\phi})$ which depends on the variable $\dot{\phi}$.  In this quantization formalism, the conjugate momentum $p_{\dot{\phi}}$ corresponds to  the
differential operator $-id/d\dot{\phi}$ and the  Hamiltonian  constraint ${\cal H} \psi(\dot{\phi}) = 0$ is replaced by the
Wheeler-DeWitt (WDW)  equation \cite{DeWitt:1967yk}, such that 
\begin{equation}
{\cal H} \psi(\dot{\phi}) = 0,\,\,\,\,\,\,\,\Rightarrow\,\,\,\,\,
\left( - \frac{1}{4}\frac{d^2}{d\dot{\phi}^2}+ V(\dot{\phi})
\right)\psi(\dot{\phi}) = 0\,.
\end{equation}
Here, we have adopted the minisuperspace approximation, which is
appropriate for our model since the Universe is considered  homogeneous, isotropic, and closed. This is possible because the system has a single dynamical degree of freedom,  namely 
 $\dot{\phi}$, to which the scale factor is related through Eq.~(\ref{Cons1a}), see Ref.~\cite{Vilenkin:1987kf}.

In this formalism,  the tunneling probability $\mathcal{P}$ associated with $\dot{\phi}$ can be quantified as 
%\cite{Landau:1991wop}
\cite{Mithani:2011en}
\begin{equation}
\mathcal{P}\sim\,e^{-2\,\mathcal{S}},\label{P1}
\end{equation}
where $\mathcal{S}$ denotes the WKB tunneling action defined as \cite{Mithani:2011en}
\begin{equation}
\mathcal{S}=\int_{\dot{\phi}_-}^{\dot{\phi}_+}\,\sqrt{V(\dot{\phi})}\,d\dot{\phi}.\label{S4}
\end{equation}
Here, Eq.~(\ref{P1}) represents the probability that the Universe tunnels from the static value $\dot{\phi}_+$ to the value $\dot{\phi}_-$ situated to the left of the infinite barrier, where the potential vanishes, see Fig.~\ref{FPP1}. As discussed in Sec.~\ref{EU}, this process does not correspond to a collapse towards $a\rightarrow0$, but to a transition from the static configuration with $a_+$ to the configuration with $a_-=a(\dot{\phi}_-)$, from which the Universe begins to evolve.

In order to obtain an analytical expression for the value of $\dot{\phi}_-$, we determine the zero of the potential by expanding the term enclosed in the square brackets of Eq.~(\ref{V1}) around $\dot{\phi}_0$, where the potential diverges, obtaining
$$
\left[ \frac{1}{\Pi_\phi^{2/3}}(A\dot{\phi} +
4B\dot{\phi}^3)^{2/3} -
\frac{\kappa}{3}\,\left(\frac{A}{2}\dot{\phi}^2 + 3B\dot{\phi}^4
+C\right) \right]\simeq \frac{1}{144}\left[\frac{A^2\kappa}{B}-48C\kappa+\frac{48}{\Pi_\phi^{2/3}}(A\sqrt{A/|B|})^{2/3}\right]
$$
\begin{equation}
+\left[\frac{A\kappa}{3}+\frac{4B}{A\Pi_\phi^{2/3}}(A\sqrt{A/|B|})^{2/3}\right]\,\left(\dot{\phi}-\sqrt{\frac{A}{12|B|}}\right)^2 + \mathcal {O}\left(\dot{\phi}-\sqrt{\frac{A}{12|B|}}\right)^3.\label{V3}
\end{equation}
Thus, from Eq.~(\ref{V3}), we find that the value of $\dot{\phi}_-$, situated to the left of the infinite barrier, becomes
\begin{equation}
\dot{\phi}_{-} =\dot{\phi}_0-R_1,\label{45}
\end{equation}
where the quantities $R_1$ and $\dot{\phi}_0$ are defined as
\begin{equation}
R_1=\sqrt{-\frac{A[48B([A\sqrt{A/|B|}]^{2/3}-\kappa C\Pi_{\phi}^{2/3})+\kappa A^2\Pi_{\phi}^{2/3}]}{48B[12B (A\sqrt{A/|B|})^{2/3}+\kappa A^2\Pi_{\phi}^{2/3}]}},\,\,\,\mbox{and}\,\,\,\,\,\dot{\phi}_0=\sqrt{\frac{A}{12|B|}}.
\end{equation}

Here, we have taken the negative branch of the root in Eq.~(\ref{45}) in order to ensure that the solution lies on the left side of the infinite barrier, where the zero of the potential is located.

In addition, from Ref.~\cite{delCampo:2011mq} the emergent value $\dot{\phi}_+$ introduced in Sec.~\ref{EU} is given by
\begin{equation}
\dot{\phi}_+=\sqrt{\frac{-A+\sqrt{A^2+12BC}}{6B}}=\dot{\phi}_0 + R_2,\label{P+}
\end{equation}
where
\begin{equation}
R_2=\dot{\phi}_+-\dot{\phi}_0=\sqrt{\frac{-A+\sqrt{A^2+12BC}}{6B}}-\sqrt{\frac{A}{12|B|}}.
\end{equation}

Considering Eqs.~(\ref{45}) and (\ref{P+}), the WKB tunneling action given by Eq.~(\ref{S4})  can be written as

\begin{equation}
\mathcal{S} = \int_{\dot{\phi}_-}^{\dot{\phi}_+}\,\sqrt{V(\dot{\phi})}\,d\dot{\phi} \approx F(\dot{\phi}_0)\int_{\dot{\phi}_0-R_1}^{\dot{\phi}_0+R_2} P\left[\frac{1}{(1 - \dot{\phi}/\dot{\phi}_0)}\right]d\dot{\phi},\label{St}
\end{equation}
where the constant $F(\dot{\phi}_0)$ is defined as
\begin{equation}
F (\dot{\phi}_0) = \sqrt{\frac{9\left[ \frac{1}{\Pi_\phi^{2/3}}(A\dot{\phi}_0 + 4B\dot{\phi}^3_0)^{8/3} - \frac{\kappa}{3}\,\left(\frac{A}{2}\dot{\phi}^2_0 + 3B\dot{\phi}^4_0
+C\right)(A\dot{\phi}_0 + 4B\dot{\phi}^3_0)^2 \right]}{A(1 + \dot{\phi}_0/\dot{\phi}_0)^2}} ,
\label{Vaprox}
\end{equation}
or equivalently 
\begin{equation}
  F (\dot{\phi}_0) =  \left(\left[\frac{\left(A \sqrt{\frac{A}{|B|}} \right)^{2/3}}{3 \Pi_\phi^{2/3}} +\frac{\kappa  \left(A^2 - 48B\,C\right)}{144 B}\right]\frac{A^2}{12 |B|}\right)^{1/2}.
\end{equation}

Although both branches of the square root are allowed by analyticity in Eq.~(\ref{St}), the semiclassical interpretation of the tunneling probability, Eq.~(\ref{P1}), requires \(\mathcal{S}>0\) in order to ensure \(\mathcal{P}\le 1\). This condition uniquely fixes the branch of the square root used in Eq.~(\ref{St}). With this choice, Eq.~(\ref{St}) follows, with \(P\) denoting the Cauchy principal value; see Refs.~\cite{Apostol:1981, Moffat:2025tpd, Guendelman:2023spd}.

In order to determine the value of the constant $\Pi_{\phi}$, we evaluate this quantity at the moment when both the scale factor and $\dot{\phi}$ take the values corresponding to the emergent Universe scenario. Thus, from Eq.~(\ref{Cons1}),  we find that the constant $\Pi_\phi$ can be written as
$$
\Pi_\phi=a_+^{3}\,[A\dot{\phi}_++4\,B\dot{\phi}_+^{3}]=12 \sqrt{6}\,\sqrt{\frac{\sqrt{A^2+12BC}-A}{B}}\,\,\,\,\times
$$
\begin{equation}
\label{Cons2}
(A+2\sqrt{A^2+12BC})\,\left[\frac{B}{(A^2+24BC-A\sqrt{A^2+12BC})\kappa}\right]^{3/2}.
\end{equation}
Here, we have utilized that the scale factor $a_+$ is given  by Eq.~(\ref{consta}) and the emergent value $\dot{\phi}_+$ is defined by Eq.~(\ref{P+}).

Thus, integrating Eq.~(\ref{St}), we  find that the WKB tunneling action  related to the probability $\mathcal{P}$ becomes
\begin{equation}
\mathcal{S} \approx F(\dot{\phi}_0)\,\dot{\phi}_0\,\log\left[\frac{R_2}{R_1}\right]\,.\label{S1}
\end{equation}
As a result, we find that   the tunneling probability exhibits a characteristic power-law behavior given by 
\begin{equation}
\mathcal{P}\sim \left[\frac{R_1}{R_2}\right]^{2\dot{\phi}_0 F(\dot{\phi}_0)}=\left[\frac{R_1}{R_2}\right]^{\gamma(\dot{\phi}_0)},\,\,\,\,\mbox{with}\,\,\,\,\gamma(\dot{\phi}_0)=2\dot{\phi}_0 F(\dot{\phi}_0),\label{P2}
\end{equation}
instead of the standard exponential expression associated to semiclassical tunneling processes. Since $R_2>R_1$ and  the exponent $\gamma(\dot{\phi}_0)$ is positive, the tunneling probability decreases as the ratio $R_2/R_1$ increases. This implies that transitions with increasingly asymmetric turning points, i.e., with increasing $R_2/R_1$, yield progressively smaller probabilities. In addition,  for large values of the exponent $\gamma(\dot{\phi}_0)\gg 1$, the tunneling probability is strongly suppressed. For the case in which the exponent   $\gamma(\dot{\phi}_0)\ll 1$, the power-law suppression becomes very weak and the probability decreases only slowly with the ratio, implying that quantum tunneling is enhanced.
In particular, using the numerical values  of the parameters given in the caption of  Fig.~\ref{Tres-Sol}, we find that the exponent $\gamma(\dot{\phi}_0)\simeq 3\times 10^{-5}\ll 1$ and  the tunneling probability $\mathcal{P}$ defined by Eq.~(\ref{P2}) is very close to  unity.

We now examine in more detail the evolution following the tunneling event. The Universe emerges with the  conditions $a(\dot{\phi}_-)=a_-$ and $\dot{a}_-=0$, see Eqs.~(\ref{Friedmann}) and (\ref{V1}).
Then, as $\dot{\phi}$ goes asymptotically to zero after tunneling, see Fig.~\ref{FPP1}, we have that $H$ goes from zero to a constant value $H_0$, where $H_0 = \sqrt{\frac{\kappa}{3}\,C}$. This process is known as the superinflation stage. During this period we have $\dot{H}>0$. We note from Eq.~(\ref{DH}) that this behavior can be achieved without violating the null energy condition $\rho+p>0$, owing to the positive spatial curvature of the closed Universe.
This is a distinctive feature of the EU scenario, which produces a suppression of the CMB anisotropies at large scales, see Ref.~\cite{Labrana:2013oca}, and after the superinflationary period we arrive at the inflationary stage.

\section{Discussion and Conclusions}\label{DC}

In this article, we have studied the quantum transition from a classically stable EU to a superinflationary phase  and  its subsequent evolution toward slow-roll inflation within the framework of TMT with spontaneously broken scale invariance. In this scenario, the Universe starts from a  classically stable Einstein static state, thereby avoiding the initial singularity and remaining in this phase until the quantum tunneling event takes place.

In contrast to the standard proposals of quantum cosmology, where quantum effects are responsible for the creation of spacetime itself, the tunneling process considered here takes place in a Universe that already exists as a classically stable EU. Quantum tunneling simply provides the mechanism that drives the transition from this static stage to a superinflationary phase, which subsequently evolves into the standard slow-roll inflationary epoch, i.e., the tunneling from $\dot{\phi}_+$ to $\dot{\phi}_-$.

In this context, we have derived an effective potential in terms of  $\dot{\phi}$,  $V(\dot{\phi})$, which  provides a description of the dynamics of the EU. Using the conserved quantity associated with the scalar field, we have expressed the scale factor as a function of $\dot{\phi}$ and rewritten the  cosmological evolution in terms of this effective potential, see Eq.~(\ref{ddf}).
This new formulation shows the existence of an infinite potential barrier that separates the static EU from the subsequent inflationary evolution. This barrier is what makes the  quantum tunneling process possible. After reconstructing the corresponding effective action in terms of $\dot{\phi}$ and $\ddot{\phi}$, 
we quantized the minisuperspace Hamiltonian, leading to 
 the Wheeler-DeWitt equation governing the quantum dynamics. 
Thus, the tunneling probability in terms of $\dot{\phi}$ was then evaluated using the WKB approximation, determining  
the probability that the Universe tunnels from the static value $\dot{\phi}_+$ to the value $\dot{\phi}_-$ situated to the left of the infinite barrier. 

In our  analysis we have found that the tunneling probability does not exhibit the conventional exponential suppression characteristic of semiclassical tunneling processes. Instead, due to the logarithmic contribution generated by the integration on the effective potential, the tunneling probability  follows a power-law behavior, see Eqs.~(\ref{S1}) and (\ref{P2}). For the parameter values considered in this work, we have obtained that the exponent associated to the power-law behavior is very small, implying that the tunneling probability is close to unity. Therefore, once the Universe reaches the emergent state, the transition to the superinflationary phase occurs with a high probability, followed by the standard slow-roll inflationary epoch. 

After the tunneling process, the Universe emerges with a finite scale factor and a vanishing   Hubble parameter, while $\dot{\phi}$ tunnels through the infinite barrier to the value $\dot{\phi}_-$, initiating the subsequent superinflationary phase. Subsequently, $\dot{\phi}$ evolves toward smaller values (to the left of the barrier), the Hubble parameter 
increases from zero to its asymptotic constant value $H_0=\sqrt{\kappa C/3}$, and the Universe evolves into the standard slow-roll inflationary regime.
The present analysis therefore provides a consistent quantum mechanism connecting the stable Einstein static Universe, quantum tunneling, superinflation, slow-roll inflation, and the subsequent standard cosmological evolution.

Several directions deserve further investigation. In particular, a detailed study  of the scalar and tensor perturbations spectra generated during the superinflationary phase following the tunneling process is left for future work. Finally, the alternative tunneling process through the finite barrier located to the right of the effective potential $V(\dot{\phi})$, towards the second classically allowed region discussed in Sec.~\ref{EU}, remains an interesting direction for future investigation, since it may lead to novel and distinct cosmological evolutions.  In particular the non linear k-essence terms present in these directions, which we have seen contribute to the vacuum energy, can be shown to contribute, when considering perturbations, to an effective Dark Matter contribution, and in this way the Dark Matter may find an explanation this way, see \cite{Guendelman:2023jsk}. 
 
\begin{acknowledgments}
E.G. is  grateful to  COSMOVERSE,  COST ACTION CA21136 and  to COST ACTION CA23130 - Bridging high and low energies in search of quantum gravity (BridgeQG),  to Ben-Gurion University of the Negev for generous support.
R.H. acknowledges the financial support of the  Agencia Nacional de Investigación y Desarrollo, ANID / Concurso FOVI-250062, 2025. P. L. was partially supported by Direcci\'on de Investigaci\'on y Creaci\'on Art\'{\i}stica de la Universidad del B\'{\i}o-B\'{\i}o through Grants RE2320212 and GI2310339. 
\end{acknowledgments}

\appendix
\section{Solutions for $\dot{\phi}$}\label{app2}

In this appendix, we show the three solutions of $\dot{\phi}$ as a function of
$a$
from Eq.~(\ref{Cons1}). The three solutions can be written as \cite{delCampo:2011mq}
\begin{eqnarray}
\dot{\phi}_1 &=& -\frac{a^3 A}{2\,3^{1/3} \left(9 a^6 B^2 \Pi_\phi
+\sqrt{3} \sqrt{a^{18} A^3 B^3+27 a^{12} B^4 \Pi_\phi
^2}\right)^{1/3}} \label{cu1}\\
\nonumber \\
&&+ \frac{\left(9 a^6 B^2 \Pi_\phi +\sqrt{3}\sqrt{a^{18} A^3 B^3+27
a^{12}B^4 \Pi_\phi^2}\right)^{1/3}}{2\,3^{2/3} a^3 B}\,,\nonumber \\
\nonumber \\
\dot{\phi}_2 &=& \frac{\left(1+i \sqrt{3}\right) a^3 A}{4 3^{1/3}
\left(9 a^6 B^2 \Pi_\phi +\sqrt{3} \sqrt{a^{18} A^3 B^3+27 a^{12}
B^4 \Pi_\phi^2}\right)^{1/3}} \label{cu2}\\
\nonumber \\
&&-\frac{\left(1-i \sqrt{3}\right) \left(9 a^6 B^2 \Pi_\phi
+\sqrt{3} \sqrt{a^{18} A^3 B^3+27 a^{12} B^4
\Pi_\phi^2}\right)^{1/3}}{4 3^{2/3}
a^3 B}\,,\nonumber \\ \nonumber\\
\dot{\phi}_3 &=& \frac{\left(1-i \sqrt{3}\right) a^3 A}{4 3^{1/3}
\left(9 a^6 B^2 \Pi_\phi +\sqrt{3} \sqrt{a^{18} A^3 B^3+27 a^{12}
B^4 \Pi_\phi^2}\right)^{1/3}}\label{cu3}\\
\nonumber \\
&&-\frac{\left(1+i \sqrt{3}\right) \left(9 a^6 B^2 \Pi_\phi
+\sqrt{3} \sqrt{a^{18} A^3 B^3+27 a^{12} B^4
\Pi_\phi^2}\right)^{1/3}}{4 3^{2/3} a^3 B}\,.\nonumber
\end{eqnarray}

\begin{figure}[h]
\begin{center}
\includegraphics[width=2.1in,angle=0,clip=true]{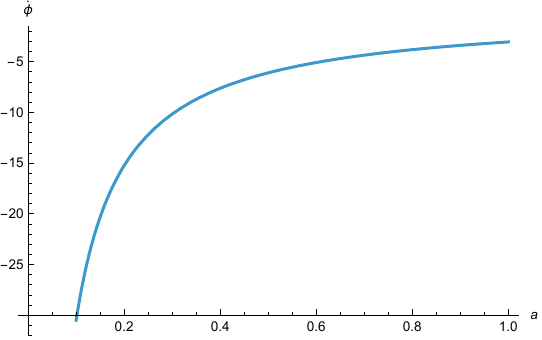}
\includegraphics[width=2.1in,angle=0,clip=true]{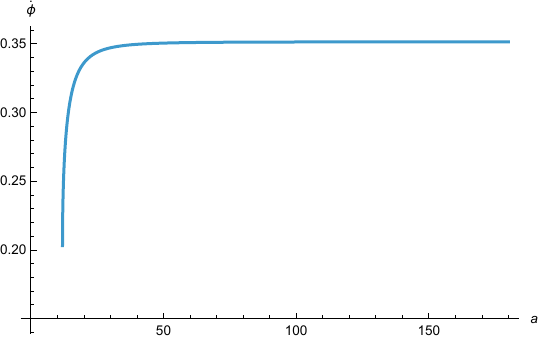}
\includegraphics[width=2.1in,angle=0,clip=true]{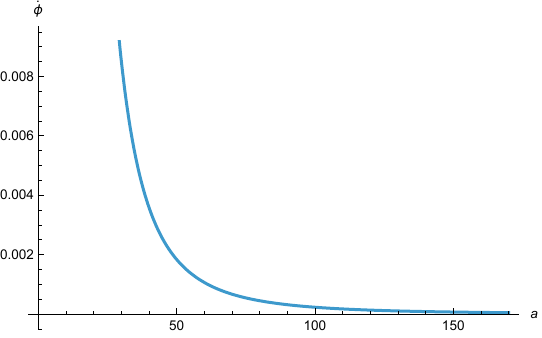}
\caption{From left to right the solutions $\dot{\phi}_1$, $\dot{\phi}_2$ and $\dot{\phi}_3$ as a function of $a$ obtained from Eq.~(\ref{Cons1}) and defined by Eqs.~(\ref{cu1}), (\ref{cu2}) and (\ref{cu3}), respectively. Here we have 
 considered the constant  $\Pi_\phi = 113.41$ together with the parameter values  of Ref.~\cite{delCampo:2011mq}.\label{F3}}
\end{center}
\end{figure}

 These solutions are shown in Fig.~\ref{F3}.
%%%%%%%%%%%%%%%%%%%%%
%%%%%%%%%%%%%%%%%%%%%%%%%%%%%%%

The solution $\dot{\phi}_1$ corresponds to the branch $\dot{\phi}<-\sqrt{A/(4|B|)}$, for which the energy density is negative and which is therefore not of physical interest, as discussed in Sec.~\ref{EU} and in Ref.~\cite{delCampo:2011mq}.

Regarding the remaining two solutions, we note from the Friedmann equation given by  Eq.~(\ref{Friedmann}) and 
Eq.~(\ref{Cons1}), which  relates the scale factor and $\dot{\phi}$, that the 
solution given by  Eq.~(\ref{cu2}) satisfies $\dot{\phi}_2 >\dot{\phi}_0$, whereas the
solution $\dot{\phi}_3$ defined by Eq.~(\ref{cu3}) satisfies $\dot{\phi}_3 <\dot{\phi}_0$, where $\dot{\phi}_0$ is given by Eq.~(\ref{45}).
Thus, from Eq.~(\ref{V1}) we observe that $\dot{\phi}_2$ and $\dot{\phi}_3$ are classically disconnected, since
$V\rightarrow \infty$ at $\dot{\phi}= \dot{\phi}_0$, so that the line defined by $\dot{\phi} = \dot{\phi}_0$ cannot be crossed classically, see Fig.~\ref{F3} where $\dot{\phi}_0=0.20$ \cite{delCampo:2011mq}. We emphasize that the tunneling process studied in this work is precisely the one connecting these two classically disconnected solutions.

\bibliography{bio}

\end{document}